\documentclass{IEEEtran} %See documentation for other class options

\usepackage{stfloats}

\usepackage{graphicx}
\usepackage{verbatim}
\usepackage{amsmath}
\usepackage{algpseudocode}
\usepackage{algorithmicx}
\usepackage{tabularx}
\usepackage{amssymb}
\usepackage{multirow}
\usepackage{hhline}
\usepackage{color}
\usepackage{listings}

\usepackage[nolist]{acronym}
\begin{acronym}
	\acro{AAA}{Authentication, Authorization and Accounting}
	\acro{AP}{Access Point}
	\acro{API}{Application Programming Interface}
	\acro{BP}{Baseline Pseudonym}
	\acro{BS}{Base Station} 
    \acro{BSM}{Basic Safety Message}
    \acro{BYOD}{Bring Your Own Device}
	\acro{C2C-CC}{Car2Car Communication Consortium}
	\acro{CA}{Certification Authority}
	\acro{CN}{Common Name}
	\acro{CAM}{Cooperative Awareness Message}
	\acro{CIA}{Confidentiality, Integrity and Availability}
	\acro{CRL}{Certificate Revocation List}
	\acro{CDN}{Content Delivery Network}
	\acro{CSR}{Certificate Signing Requests}
	\acro{DAA}{Direct Anonymous Attestation}
	\acro{DDoS}{Distributed Denial of Service}
	\acro{DDH}{Decisional Diffie-Helman}
	\acro{DENM}{Decentralized Environmental Notification Message}
	\acro{DHT}{Distributed Hash Table}
	\acro{DoS}{Denial of Service}
	\acro{DoT}{Department of Transportation}
	\acro{DPA}{Data Protection Agency}
	\acro{DTN}{Delay Tolerant Network}
	\acro{ECU}{Electronic Control Unit}
	\acro{ETSI}{European Telecommunications Standards Institute}
	\acro{ECDSA}{Elliptic Curve Digital Signature Algorithm}
	\acro{ECC}{Elliptic Curve Cryptography}
	\acro{EVITA}{E-safety Vehicle Intrusion protected Applications}
	\acro{FCFS}{First-Come First-Served}
	\acro{FOT}{Field Operational Test}
	\acro{FPGA}{Field-Programmable Gate Array}
	\acro{GN}{GeoNetworking}
	\acro{GS}{Group Signatures}
	\acro{GM}{Group Manager}
	\acro{GBA}{Generic Bootstrapping Architecture}
	\acro{GPA}{Global Passive Adversary}
	\acro{GUI}{Graphic User Interface}
	\acro{HP}{Hybrid Pseudonym}
	\acro{HSM}{Hardware Security Module}
	\acro{HTTP}{Hypertext Transfer Protocol}
	\acro{IEEE}{Institute of Electrical and Electronics Engineers}
	\acro{IoT}{Internet of Things}
	\acro{ITS}{Intelligent Transport System}
	\acro{IT}{Information Technologies}
	\acro{IMSI}{International Mobile Subscriber Identity}
	\acro{IMEI}{International Mobile Station Equipment Identity}
	\acro{IdP}{Identity Provider}
	\acro{ISP}{Internet Service Provider}
	\acro{LEA}{Law Enforcement Agency}
	\acro{LTC}{Long-Term Certificate}
	\acro{LTCA}{Long-Term \acl{CA}}
	\acro{LDAP}{Lightweight Directory Access Protocol}
	\acro{LTE}{Long Term Evolution}
	\acro{LBS}{Location-based Service}
	\acro{MAC}{Message Authentication Code}
	\acro{MCA}{Message \ac{CA}}
	\acro{MEA}{Misbehavior Evaluation Authority}
	\acro{MSN}{Mobile Social Network}

	\acro{OBU}{On-Board Unit}
	\acro{OCSP}{Online Certificate Status Protocol}
	\acro{OSN}{Online Social Network}
	\acro{PC}{Pseudonymous Certificate}
	\acro{PCA}{Pseudonymous \acl{CA}}
	\acro{PDP}{Policy Decision Point}
	\acro{PEP}{Policy Enforcement Point}
	\acro{PIR}{Private Information Retrieval}
	\acro{PKI}{Public-Key Infrastructure}
	\acro{POI}{Point of Interest}
	\acro{PRECIOSA}{Privacy Enabled Capability in Co-operative Systems and Safety Applications}
	\acro{PRESERVE}{Preparing Secure Vehicle-to-X Communication Systems}
	\acro{PS}{Participatory Sensing}
	\acro{P2P}{peer-to-peer}
	\acro{RA}{Resolution Authority}
	\acro{REST}{Representational State Transfer}
	\acro{RBAC}{Role Based Access Control}
	\acro{RCA}{Root \acl{CA}}
	\acro{RSU}{Road-Side Unit}
	\acro{SAML}{Security Assertion Markup Language}
	\acro{SCORE@F}{Système COopératif Routier Expérimental Français}
	\acro{SeVeCom}{Secure Vehicle Communication}
	\acro{SIS}{Security Infrastructure Server}
	\acro{SP}{Service Provider}
	\acro{SSO}{Single-Sign-On}
	\acro{SoA}{Service-oriented-Approach}
	\acro{SOAP}{Simple Object Access Protocol}
	\acro{SAS}{Sample Aggregation Service}
	\acro{TS}{Task Service}
	\acro{TLS}{Transport Layer Security}
	\acro{TPM}{Trusted Platform Module}
	\acro{TTP}{Trusted Third Party}
	\acro{TVR}{Ticket Validation Repository}
	\acro{URI}{Uniform Resource Identifier}
	\acro{VANET}{Vehicular Ad-hoc Network}
	\acro{V2I}{Vehicle-to-Infrastructure}
	\acro{V2V}{Vehicle-to-Vehicle}
	\acro{V2X}{\ac{V2V} and/or \ac{V2I}}
	\acro{VC}{Vehicular Communication}
	\acro{VM}{Virtual Machine}
	\acro{VSN}{Vehicular Social Network}
	\acro{VSS}{\ac{VC} Security Subsystem}
	\acro{WAVE}{Wireless Access in Vehicular Environments}
	\acro{WSDL}{Web Services Discovery Language}
	\acro{W3C}{World Wide Web Consortium}
	\acro{VANET}{Vehicular Ad-hoc Network}
	\acro{VPKI}{Vehicular Public-Key Infrastructure}
	\acro{WS}{Web Service}
	\acro{WoT}{Web of Trust}
	\acro{WSACA}{\ac{WAVE} Service Advertisement \ac{CA}}
	\acro{XML}{Extensible Markup Language}
	\acro{XACML}{eXtensible Access Control Markup Language}
	\acro{3G}{3rd Generation}
\end{acronym}

\usepackage{fixltx2e,fix-cm}
\usepackage{amssymb}
\usepackage{amsmath}
\usepackage{graphicx}
\usepackage{mathptmx}
\usepackage{comment}

\begin{document}
	
\title{Security and Privacy in Vehicular Social Networks}

\author{Hongyu Jin, Mohammad Khodaei and Panos Papadimitratos\\Networked Systems Security Group, KTH Royal Institute of Technology, Sweden \\
		\emph{\{hongyuj, khodaei, papadim\}}@kth.se\\
		www.eecs.kth.se/nss}

% use for special paper notices
%\IEEEspecialpapernotice{(Invited Paper)}

% make the title area
\maketitle

%\frontmatter

%\title{Saunders Template} %This is a placeholder titlepage, it will not be final.
%\author{Yours Truly}
%\maketitle

%\include{frontmatter/dedication}
%\cleardoublepage
%\setcounter{page}{7} %previous pages will be reserved for frontmatter to be added in later.
%\tableofcontents
%\include{frontmatter/foreword}
%\include{frontmatter/preface}
%\listoffigures
%\listoftables
%\include{frontmatter/contributor}
%\include{frontmatter/symbollist}

%\mainmatter

\section{Introduction}
\label{sec:intro}

During the past decade, the trend is to enable vehicle communication, by equipping them with \acp{OBU}. \acf{VC} systems facilitate \acp{ITS}, enabling various applications on top of \ac{V2V} and \ac{V2I} communication. Vehicles can also access various \acp{SP} through \acp{BS}, \acp{AP} and \acp{RSU}. Basically, original \ac{VC} applications aim at providing awareness to avoid vehicle collisions and helping drivers choose better routes based on traffic density \cite{papadimitratos2007architecture}. This is achieved by vehicles' active periodical beaconing of the current status and sensed context information (e.g., obstacles or accidents). These transmissions leverage \acp{OBU} pre-installed in the vehicles and do not assume any relationships with neighbors before the transmissions and receptions.

% Beyond such uni-directional transmission/broadcast, the communication capability lay the foundation for various interactions among the vehicles.

At the same time, interconnected vehicles facilitate message exchange beyond transportation safety and efficiency. This enables socializing with the drivers and passengers of nearby vehicles. Unlike \acp{OSN} and most \acp{MSN}, the users/devices (nodes) in \acfp{VSN} mostly interact when they are within communication range (i.e., physically close to each other, as determined by their trips). Due to the mobility of vehicles, they have ephemeral encounters and interactions. However, vehicle interactions could exploit such characteristics and even promote content dissemination in \acp{VSN} thanks to broad network of vehicle contacts during the trips. Moreover, vehicle connectivity to the Internet (leveraging \acp{BS} or \acp{AP}) can enable interactions among \acp{VSN}, \acp{OSN} and \acp{MSN}.

Traditional social networks leverage long-term user identities (i.e., an identity is created based on, e.g., an e-mail address or a username, and cannot be changed during its lifespan), and all user activities are carried out under these identities. The users also interact based on established relationships (e.g., within friends or group members), which are linked to their identities. However, these identities do not necessarily indicate the true identities of the users (i.e., user profiles can be faked); thus, no strong authentication is required: the users are not kept accountable for their actions. On the contrary, in \ac{VC} systems, there is consensus in academia and industry that vehicles do not expose long-term identities due to privacy concerns, rather, short-term and unlinkable identities should be used to preserve user privacy. While privacy is important in \ac{VC} systems, strong identification of drivers and vehicles is needed considering the high stakes in traffic systems (notably, driver and passenger safety): the messages in \ac{VC} systems need to be properly protected by proving that the messages are originated from legitimate users in the system, in order to guarantee the secure and privacy preserving operations in the system. Both of the requirements can be achieved leveraging pseudonymous authentication~\cite{papadimitratos2006securing, leinmueller2006sevecom, c2c, 1609draft, papadimitratos2007architecture,gisdakis2013serosa, khodaei2014towards}. In fact, security and privacy in \ac{VC} systems have been extensively studied and significant effort has already been made towards the deployment of secure \ac{VC} systems, which are the basis for secure \acp{ITS}.

% : each vehicle is provided with a set of pseudonyms (authenticated public keys by the Certification Authorities (CAs))~\cite{papadimitratos2008secure}, and the vehicle switches from one pseudonym to another over time to increase the unlinkability of user activities.
% Interactions in \acp{VSN} are usually carried out among the vehicles physically close to each other: the vehicles that are involved in radio ranges of each other could maintain temporary social relationships.

\acp{VSN} consider the \ac{VC} network as the underlying networking facility, along with its location and context-specific services and features. While we embrace emerging \ac{VSN} applications, it is important that \ac{VC} system security and privacy are not compromised by the \ac{VSN} functionality. \acp{VSN} could and in fact should build upon the security infrastructures designed and deployed for \ac{VC} systems and seek to address \ac{VSN}-specific requirements based on extensions or tailoring of those security infrastructures. Moreover, security solutions proposed for the relevant areas (e.g., \ac{LBS} and participatory sensing) could evolve and be integrated into \acp{VSN}. This could largely promote the popularity and deployment of \acp{VSN} rather than building the whole infrastructure from scratch; this is what we advocate in this chapter. We outline the \ac{VSN} architecture and content dissemination in different architectures. We continue with the investigation of security and privacy requirements in the \ac{VC} landscape. This is important so that \ac{VSN} can be deployed, possibly promising the adoption of \ac{VC} technology itself, while ensuring the strong security and privacy protection for the overall system. Moreover, we survey the existing security and privacy solutions for emerging applications (which are potential applications for \acp{VSN}) and show that they could be integrated to the \acp{VSN} eliminating the need to introduce redundant components to the system. We close this chapter with a discussion of open challenges for the security and privacy, and a brief conclusion.

\section{Vehicular Social Networks} 
\label{sec:vsn}

\acp{OSN} with rich features have been integrated into people's daily lives. They have satisfied users' demand on socializing with friends or making new friends among people with common interests. Nowadays, \acp{OSN} are easily accessible from mobile devices (e.g., smartphones) and many of them exploit user mobility, thus, they are location-aware. Users can look for nearby users or tag posts with their current locations; this way they can be discovered by other users with location-based searching.

\acp{OSN} maintain steady user relationships: users with  common interests have direct or indirect relationships. Leveraging the Internet, user interactions are not time-/space-restricted. Although there exist decentralized social networks (e.g., Synereo\footnote{www.synereo.com}), the dominant \acp{OSN} (e.g., facebook\footnote{www.facebook.com} and twitter
\footnote{www.twitter.com}) are centralized, with their servers storing information related to the users or the data generated and disseminated by the users. Most \acp{OSN} follow a publisher/subscriber model: users publish the content to the central server and the central server disseminates the data to the users who have subscribed to the content (e.g., followers or friends). The content dissemination in \acp{OSN} is not necessarily a real-time process: the users can see the content at any time they wish as long as they have access to the Internet.

Social networks can also be decentralized. Decentralized social networks highlight users' control over their own data. The data are stored locally and shared with other users who they trust or closely relate to. Leveraging decentralization of social networks, user mobility could be exploited in \acp{MSN} to promote information sharing and region-specific interactions. Decentralized social networks emerged mainly due to privacy concerns in centralized \acp{OSN}~\cite{verma2013privacy, cutillo2009safebook, mezzour2009privacy}. Centralized servers could breach user privacy simply because all user-related sensitive data are stored in those central servers: data are exposed once the server is compromised or even the central server itself could be interested in the data. This coincides with privacy concerns in the context of \acp{VSN}, as it will become clear in the discussion below.

\begin{figure}[!ht]
	\centering
	\includegraphics[width=\columnwidth]{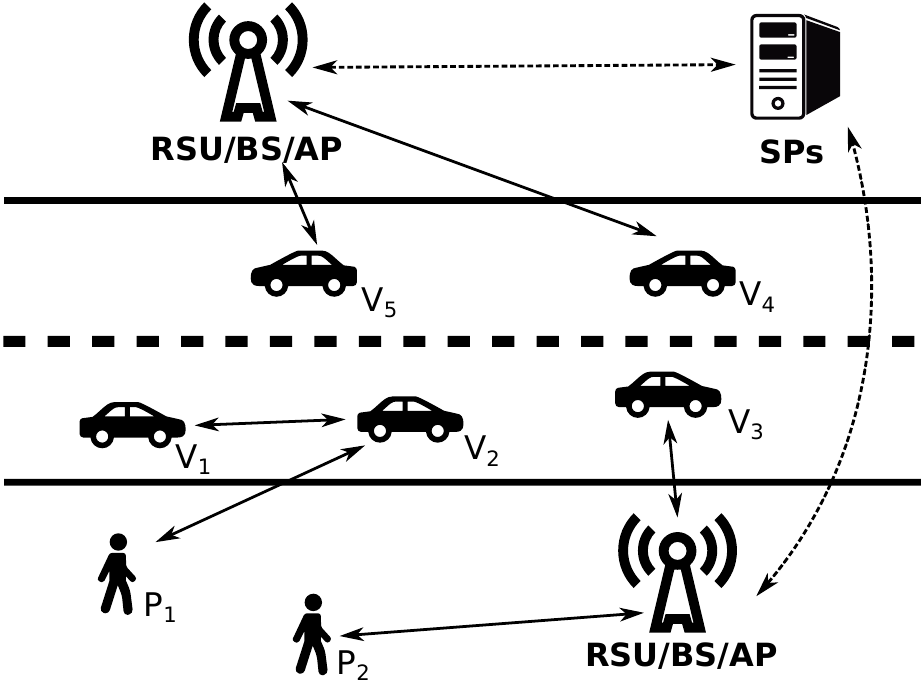}
	\caption{Illustration of \acp{VSN}: (1) Vehicles (with \acp{OBU}, e.g., $V_3$, $V_4$ and $V_5$) and users (with smartphones, e.g., $P_2$) can access various \acfp{SP} via \acfp{RSU}, \acfp{BS} or \acfp{AP}; (2) Vehicles (e.g., $V_1$ and $V_2$) or users (e.g., $P_1$) can interact with each other in an ad-hoc network (e.g., share information obtained from \acp{SP}). (Icons made by Freepik from www.flaticon.com)}
	\label{fig:vsn}
\end{figure}

Essentially, user socialization could appear in any network where communication (thus user interactions) is convenient. As described earlier, vehicles are communication-enabled thanks to the \acp{OBU} pre-installed in the vehicles. Thus, interconnected vehicles enable drivers and passengers to socialize with other nearby users, forming \acp{VSN}. \acp{VSN} inherit characteristics of traditional social networks but they also have their own properties. In principle, \acp{VSN} are social networks built on top of \acp{VANET} and considered as an extension of user-centric social networks. Applications in the \acp{VSN} could be based on the purpose of \ac{VC} systems, e.g., safety applications, while entertainment applications could also be involved. We discuss different characteristics of \acp{VSN} illustrated in Fig.~\ref{fig:vsn} in the rest of this section. 

%(i.e., the network of vehicles)

\subsection{Networking Architecture}

An \ac{OBU} could integrate, an IEEE 802.11p interface as well as cellular and Wi-Fi interfaces. Through the IEEE 802.11p interface, vehicles communicate with other vehicles (\ac{V2V}) or with \acp{RSU} (\ac{V2I}). Cellular and Wi-Fi interfaces enable connection to the Internet via \acp{BS} and \acp{AP}, and access to various \acp{SP}.

\acp{VSN} can be either centralized, decentralized or in a hybrid form. Similar to \acp{OSN}, a centralized \ac{VSN} involves a central server, and users can interact via the central server. Decentralized \acp{VSN} could leverage \acp{VANET} to form groups on-the-fly and enable communication among the group members. In addition, information obtained from the central servers could be shared with other nodes in an ad hoc manner.

%Centralized \acp{VSN} could leverage \acp{RSU} relaying the content to the central servers. Moreover, mobile devices that have cellular data interfaces could simply connect to the Internet via \acp{BS}.

%Decentralized \acp{VSN} lack a centralized infrastructure to maintain and monitor activities of vehicles as most of the communication are carried out in an ad-hoc manner. Vehicles could share the data they obtained from the content providers with other vehicles.

% This scales up the system while the user privacy is protected~\cite{shokri2014hiding}. The lack of central authorities or monitoring facility implies that security is at stake, since any vehicle (or mobile device) could transmit anything they want in the network, which pose security threats to other participants.

\subsection{Participation and Social Relations}

%participate in the \acp{VSN} and 
The participants in the \acp{VSN} are not limited to vehicles equipped with \acp{OBU}, but they can also be passengers and pedestrians using smartphones. The \acp{VSN} leverage \ac{VC} systems while highlighting the social connections between the participants. Smartphone users could bring \acp{OSN} and \acp{VSN} closer. In fact, it becomes more common that data are shared among applications. \ac{VSN} users could share data they obtained from the \ac{VSN} or even the \ac{VC} system within other \acp{OSN} they join. Such interactions among the \acp{OSN} and \acp{VSN} could promote the popularization of the \ac{VSN} applications. Moreover, smartphone-based \acp{ITS}~\cite{manolopoulos2011securing,zaldivar2011providing,fazeen2012safe,gisdakis2014secure} leverage smartphones and have been proposed as an alternative approach before \ac{OBU}-equipped vehicles become universal.

\ac{VSN} applications could exploit different kinds of user relations. Interest-based applications could maintain long-term relations among users. \ac{VSN} users could be friends in \ac{OSN} applications and at the same time interact with \ac{VC}-related and \ac{ITS}-related information within the \ac{VSN} context, e.g., wish to share traffic-related information. In this case, \ac{OSN} application only provide a way to establish relation in \ac{VSN} applications. \ac{VSN} interactions could also strictly specific to geographic regions while these interactions are short-term and do not assume any relationships a priori. For example, passengers with same destination in public transportation can share \ac{POI} information around the destination.

%while wish to interact via \acp{VSN} and share 
%be carried out among the vehicles in a specific region 

\subsection{Applications}

Any transportation-related information could be facilitated by \ac{VSN} applications. Users can obtain traffic information from central providers and share these information with nearby users, or they can sense surrounding traffic condition and construct a global view based on contributed sensing data from multiple nearby users. This can be seen as an extension of \ac{VC} applications. \ac{LBS} is another type of important application in \ac{VSN}. Traditional \acp{LBS} (e.g., querying \acp{POI} from \ac{LBS} servers) could still exist in \acp{VSN}. However, as socialization is highlighted in \acp{VSN}, users could largely exploit their mobility and even generate customized location-dependent information (e.g., travel guide of a certain place) based on their interests, which can be later shared with other users who have similar interests. This is more dynamic and more user-centric compared to information that could be obtained from traditional \acp{LBS}. In general, users can in fact share any information they are interested in within \acp{VSN}, as long as the senders and the receivers are allowed to do so (e.g., not a copyrighted music).

% Compared to traditional social networks, no background information is assumed in \acp{VSN}. Information shared in \ac{VSN} applications could be region-based or interest-based. Content can be simply broadcasted to a specific region assuming the vehicles in that region might be interested in the information (e.g, traffic condition and \acp{POI} around). 

%\input{Chapters/chapter1/Sections/application}

\section{Security and Privacy Considerations in \acsp{VSN}}
\label{sec:security-privacy-consideration}

Security and privacy are key factors for designing and deploying a large scale trustworthy \ac{VSN}. As described earlier, \acp{VSN} are built on top of \ac{VC} systems and \ac{VSN} applications should not deteriorate achieved \ac{VC} system security and privacy. Security and privacy requirements could vary depending on \ac{VSN} applications. For example, for a safety application (e.g., hazard warning), integrity, non-repudiation and accountability are of paramount importance (unlike confidentiality) while for a traffic management application, not only the integrity but also the verifiability of the content is crucial to prevent users being misled. On the contrary, for entertainment applications, the availability of the service is important. Next, we list and explain the basic security and privacy requirements for \acp{VSN} based on those for \ac{VC} systems~\cite{papadimitratos2006securing}; while in the following subsections, we further explain security and privacy concerns for the \ac{VSN} applications.

\subsection{Basic Security and Privacy Requirements}

%In most cases, a
\textbf{Authentication and Integrity:} A node should authenticate the source of a message so that only the information from trusted, i.e., legitimate, source should be accepted. Moreover, messages should not be tampered: unauthorized entities should not be able to alter the content of the messages.

%once the message is authenticated and sent out

%that can be obtained by any vehicle 
\textbf{Confidentiality:} Information exchanged by the users could be done in confidential manner: information is accessible only by authorized recipients, e.g., vehicles in a platoon, or vehicles from the same manufacturer. Information could be simply broadcasted, which does not need to be confidential (e.g., traffic conditions disseminated to a specific region).

\textbf{Accountability and Non-repudiation:} Entities in the system, including vehicles (i.e., \acp{OBU}), smartphones and infrastructures, should be accountable for the actions they perform in the system, and should not be able to deny the actions they have performed in the system.

\textbf{Unlinkability and Anonymity:} User identities should not be exposed, i.e., users should be anonymous and their (authenticated) messages should not be linkable. However, for practicality and efficiency, we inherit conditional anonymity (i.e., pseudonymity) from the \ac{VC} domain: user messages are only linkable over a system defined period $\tau$, and users are pseudonymous as long as they do not misbehave in the system. Moreover, users should be able to gain and accumulate reputation or credits for their contribution to the system while using pseudonyms as their legitimate identities in the system.

\textbf{Access Control:} Only legitimate entities, registered within the system, should be able to operate and contribute to the system. The system should prevent any illegitimate entity from participating in system operations, e.g., content delivery or crowdsourcing. In \acp{VSN}, user interactions could also be restricted by relationships: unlink message broadcast and sender authentication in \ac{VC} systems, user interactions could be allowed strictly based on relationships (e.g., among friends).

\textbf{Availability:} The system should remain operational even in case of a fault. Especially, the functionality of underlying network architecture (i.e., user safety and traffic efficiency) should not be affected due to the system failure.

%Safety-related messages in safety applications should be disseminated in a real-time manner to be operational. The delay overhead introduced to ensure the security and privacy requirements (e.g., extra communication delay due to security payloads, and message authentication and verification processes) should be thoroughly investigated \cite{calandriello2011performance}.

%This is made a concern especially due to the delay introduced by ensuring security and privacy requirements (e.g., extra communication delay due to security payloads, and message authentication and verification processes).

%such messages could not tolerate long delay from they are being sent and until they are received and accepted.

\subsection{Adversarial Model}

\textbf{Honest-but-Curious Entities:} Recent experience from mobile applications (e.g., \acp{LBS}) \cite{nsa} shows that service providers are aggressively collecting user information in order to profile users. For example, an \ac{LBS} server could collect user queries (including user locations and interests) in order to offer customized services or push advertisements to the users. This led to the concerns from the users on their private information. In a general sense, this applies to every entity within the system, e.g., passive observers, service providers and security infrastructure entities, which can infer information in order to infringe user privacy. Many works try to solve this problem by transferring the trust to an introduced \ac{TTP}~\cite{gedik2008protecting,mokbel2006new}: a proxy is introduced between the users and the honest-but-curious server, so that all user requests are anonymized by the proxy before forwarding to the servers. However, the same concern should be applied to any entity that is introduced to solve this problem, for which those works assumed to be fully trustworthy. Essentially, if the same information is available to the servers and the introduced \acp{TTP}, then there is no difference between what the servers and those entities could do (i.e., the information they can infer).

This is why we need to extend our adversarial model from \emph{fully-trustworthy} to \emph{honest-but-curious} servers. Honest-but-curious entities never deviate from system security policies or protocols, but they are tempted to infer and exploit user sensitive information, e.g., profile users and push advertisements to users based on their interests.

\textbf{Malicious Participants:} Due to the dynamic nature (intensified in a decentralized architecture) of \acp{VSN}, registered vehicles and users (legitimate insiders) are able to disseminate faulty information to affect a process, e.g, temperature measurement. In addition, internal adversaries might try to pollute the content achieved from the content provider before they share with other users. This is due to the openness of sharing data in \acp{VSN} which leads to additional vulnerabilities than in traditional social networks. Polluted data reported from faulty insiders should be filtered out and malicious users should be evicted from the system. This requires that the accountability of user actions in the network be preserved. The situation is even worse if a malicious user is able to equip with multiple valid (yet fake) identities and affect the system with those identities. For example, an adversary could clone an identity (which he/she should not own) to mislead other users by disseminating aggressively the false information. This type of attack is well known as Sybil-based misbehavior \cite{douceur2002sybil} in which an attacker is able to clone an identity, thus creating socialbots. They can perform various kinds of attacks, e.g., injecting bogus messages to control the outcome of a specific protocol, or disseminating spams to other users.

On the contrary, external adversaries have limited capabilities to destroy the system; however, they can try to harm user privacy by eavesdropping the communication, or they could simply launch jamming and \ac{DDoS} attacks on a specific target or area to breach the system availability.

% In \acp{OSN}, most of the participants are vulnerable to several security and privacy threats \cite{fire2014online}. Many of the existing threats (e.g., malware, phishing attacks) in \acp{OSN} are eradicated thanks to the basic security requirements. However, some threats exist even in the presence of achieved basic security and privacy properties. 

\textbf{Selfish Participants:} Crowdsourcing based mobile applications \cite{froehlich2009ubigreen, smaldone2011cyber, ganti2010greengps, maia2012hydi} have been widely used for enhancing transportation efficiency and safety. These applications rely on users' participation and contribution to measure specific phenomena (e.g., temperature and traffic status). However, such applications would not work without active participation of users. In a \ac{VSN}, selfish users could try to achieve higher and optimal awards by sacrificing the minimum resources. These misbehaving internal adversaries utilize the resource of other nodes to achieve a better service without participating in the tasks \cite{kapadia2009opportunistic}. The success of these applications depends on the participation of the majority of users and their collaboration to achieve desired goals. Unless the mechanisms that motivates the user participation are in place, selfish users would not be willing to consume resources for other users or the system. Appropriate mechanisms should be provided to monitor user activities or incentivize users for their contribution to the system; the system should be able to identify selfish users or free riders, thus eradicate them from the system or degrade and limit their access to (the services in) the system.

\section{Existing Security and Privacy Solutions}
\label{sec:existing-solutions}

A lot of research effort on security and privacy have been carried out in the relevant areas, e.g., \ac{VC}, \ac{MSN} and crowdsourcing. Security and privacy solutions in those areas could be evolved and integrated to \acp{VSN} in order to address similar problems that exist in the \acp{VSN}. In this section, we introduce existing solutions from other domains that could facilitate secure and privacy preserving \acp{VSN}.

\subsection{Decentralization}

Decentralization of a system could be due to various reasons, among which privacy is one of the main concerns that motivates the decentralization. It has been considered in many works that the central servers in \acp{OSN} or \acp{LBS} are tend to collect sensitive information of users and even inferring extra information from the collected data~\cite{shokri2014hiding, gisdakis2014sppear, jin2015resilient}. Such central servers fall into honest-but-curious model. Location privacy is a main concern in \acp{VSN}, since the interactions among the entities in \acp{VSN} are location-dependent: information obtained from the servers are customized based on the geographical information of the vehicles (e.g., \acp{LBS}). The geographical information could be used to track the users and even the interests of a specific user can be inferred from the information being requested. \emph{k-anonymity}~\cite{ghinita2007mobihide, gedik2008protecting} has been widely used for protecting location information of users in both centralized or decentralized manners. Anonymizer-based approaches leverage an anonymizer introduced between the users and the servers~\cite{gedik2008protecting,mokbel2006new}. However, it has been considered in many works that such anonymizer could also be a threat for user privacy, i.e., they can also be honest-but-curious~\cite{shokri2014hiding, jin2015resilient}. Decentralized approaches have been proposed to eliminate such concerns. Users could leverage peers around them to form a region that involves $k-1$ other users and use this obfuscated region instead of an accurate location~\cite{ghinita2007mobihide, gkoulalas2010providing}. Such approaches trade off high burden on the users' mobile devices for searching nearby peers. Moreover, it is also an issue that have been pointed out in many research, the strategy of peer selections determines the efficiency and effectiveness of the schemes~\cite{ghinita2007mobihide, gkoulalas2010providing}. Especially, when node mobilities are not predictable, this would be even harder.

Such collaboration in \acp{VSN} could be made easier by forming groups in \ac{VANET} leveraging similar mobilities of vehicles. In~\cite{sampigethaya2007amoeba}, nearby vehicles form groups and maintain the groups as long as the vehicles are within each other's communication range. Each group has a leader which acts as a temporary anonymizer for the group. The group leader is rotated over time within the group in order to share the burden among the group members and limit the information that could be learned by the group leader. Such temporary centralization leverages the characteristics of \ac{VANET} and decrease the effort for searching for the most suitable peers. However, these approaches would not help if the honest-but-curious server is only interested in the symbolic locations (e.g., church, shopping mall and railway station) of the users, since all the $k$ members are likely to fall in a same symbolic location.

Content-sharing can further protect privacy of users, since the users do not need to query the content-provider for every request, from which the content-provider could learn sensitive data of the users. For example, information sharing in \ac{LBS}~\cite{shokri2014hiding} help users to protect their privacy in a collaborative way. In~\cite{shokri2014hiding}, users share \ac{LBS}-obtained information with their neighbors so that the users who need the same information does not need to query the \ac{LBS} server again. This decreases the user exposure to the \ac{LBS} server. However, it can also allow internal attackers to provide faulty information to benign users, while the receivers do not have clue if the information is valid or not, as long as they do not query the \ac{LBS} server directly. As a common issue in an open decentralized network architectures, it is vulnerable to active malicious nodes within the network. Thus, the user authentication is needed to eliminate illegitimate users from the network.

%ss  And it fits VANET.....

% Apart from the traditional \acp{OSN} applications which could be inherited by \acp{VSN}, \acp{LBS} would be one of the main components in \acp{VSN}, in which users obtain location-dependent information (e.g., \ac{POI} information) from \ac{LBS} servers.

\subsection{Pseudonymous Authentication}

To address the concern brought by the openness of decentralized systems, transmissions in \acp{VSN} should be verifiable in terms of trust, especially for the safety-related applications. In most of the \ac{VSN} applications, users are strangers to each other and had no social interaction before. Vehicles have limited time to share information with each other due to the mobility of vehicles, i.e., the \ac{V2V} communications are highly dynamic and unreliable. This implies that the users cannot leverage accumulated reputation for trust establishment among the users. \ac{PKI}-based solutions could be used to ensure authenticity and integrity of the transmitted messages, in which trust among the vehicles are established leveraging a \ac{TTP} (i.e., a \ac{CA}). However, with traditional certificate-based authentication, one can easily trace the messages related to a specific vehicle based on its identity (in the certificate), thus profile its behavior/action, especially considering the openness of wireless networks. Encryption of messages would help so that only the targeted recipients could decrypt the messages. However, as described earlier, vehicles have ephemeral encounter events so that it is unrealistic to negotiate (multiple pairs of) security associations within short period with (multiple) recipient(s) and encrypt all the transmissions. Moreover, it is hard to decide in advance the interested recipients in case the transmission are region-based/targeted, i.e., the messages should be authenticated and broadcasted to all the neighboring nodes. Thus, the approaches relying on long-term identity cannot be used since all the user actions could be linkable. This motivated many works with their solutions leveraging anonymous credentials to satisfy both the security and privacy requirements in \ac{VC} domain.

\begin{figure}[!ht]
	\centering
	\includegraphics[width=\columnwidth]{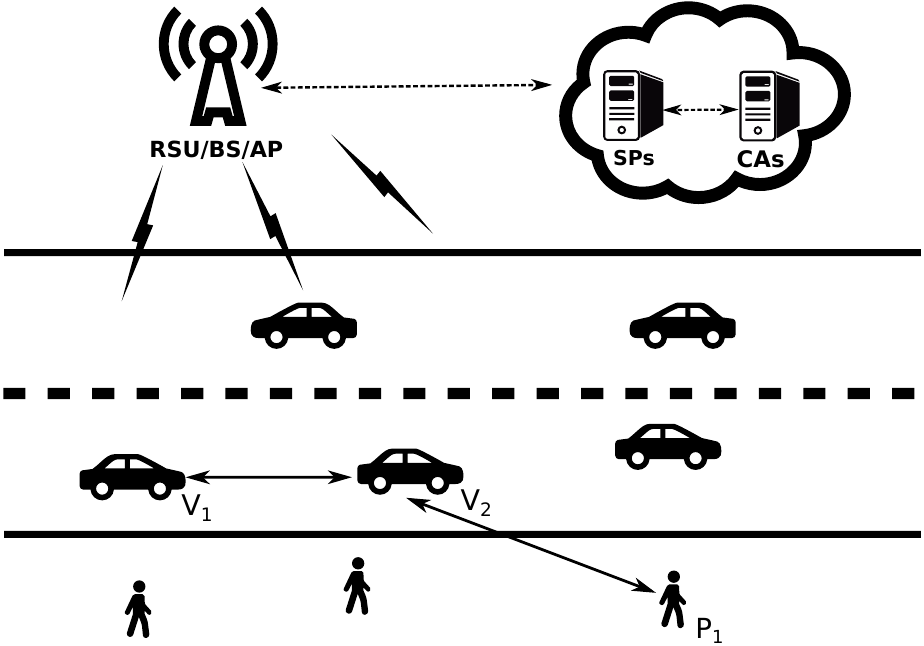}
	\caption{Vehicles and users can obtain pseudonyms from the Certification Authorities (CAs). The communication in the \acp{VSN} is protected with pseudonymous authentication including \ac{P2P} communication (e.g., $V_1$-$V_2$ and $V_2$-$P_1$) in the ad-hoc network and vehicle/user-SP communication. (Icons made by Freepik from www.flaticon.com)}
	\label{fig:psnym_auth}
\end{figure}

Generally, there are two categories of \ac{VPKI} schemes proposed for the \ac{VC} systems: public key based and group signature based schemes. The public key based schemes \cite{alexiou2013vespa, gisdakis2013serosa, studer2009tacking, schaub2010v, bismeyer2013copra, khodaei2014towards} equip users with a set of short-term (pseudonymous) credentials (i.e., pseudonyms), switching from one pseudonym to another over time. A pseudonym is a public key authenticated by the \ac{PCA}. The pseudonyms are essentially unlinkable, i.e., one cannot link two pseudonyms since they are anonymized (i.e., do not include any information that could be linkable). Each user signs an outgoing content with time- and geo-stamped using the private key corresponding to the current valid pseudonym. The content is attached with the pseudonym (and possibly the chain of trust) to facilitate the verification by the recipient. Having received a content, depending on the spatial, temporal, and interest scope of the receivers, they verify the attached pseudonyms first and then validate the signature on the content using the public key of the attached pseudonym.\footnote{We assume that the sender and receiver trust the pseudonym issuer, the PCA.} Using the anonymous pseudonyms, one can achieve integrity, non-repudiation, accountability and conditional anonymity. Pseudonymous can be integrated with different services (and their \acp{SP}) to provide secure and privacy preserving \acp{VSN}. For example, in~\cite{jin2015resilient}, a secure and privacy-enhancing \ac{LBS} is proposed leveraging information sharing and the pseudonymous authentication. Users authenticate their queries and responses under the pseudonyms obtained from the \ac{PCA}. In this way, illegitimate users are prevented from providing false information to the benign users, while internal adversaries are kept accountable for their actions. Fig.~\ref{fig:psnym_auth} shows the secure and privacy preserving \acp{VSN} architecture leveraging pseudonymous authentication.

The other schemes propose the use of group based signature schemes \cite{calandriello2007efficient, studer2009tacking, lin2007gsis, lu2008ecpp} for identity and credential management in \ac{VC} systems. These approaches leverage group signatures \cite{chaum1991group,boneh2004group,boneh2004short} in which the receiver can verify that a legitimate group member has signed a message without knowing who the signer is. In case of a misbehavior, the group manager is able to \emph{open} a signature, thus disclosing and revoking the signer's identity.

%There are different group signature schemes, e.g., bilinear pairing based cryptography \cite{boneh2001identity, shamir1985identity, chaum1991group, boneh2004group}.

While group signatures can be used to protect the transmitted messages, such schemes exhibit high computational overhead for signature generation and verification~\cite{calandriello2011performance}. The integration of group signatures and pseudonyms could decrease the computational overhead for resource-constrained devices. \cite{calandriello2011performance} proposes a hybrid approach in which the vehicles generate public/private key pairs and sign the public keys with their own group signing keys. In this way, a pseudonym needs to be verified only once for the first time it is received and cached locally during its lifetime. For the following messages signed under the cached pseudonyms, only the signatures on the messages need to be verified with the public key of the pseudonym, which are much cheaper than group signature verifications.

\subsection{Sybil Resilience}

Using such pseudonymous authentication approaches, it is possible that a compromised vehicle equipping itself with multiple simultaneously valid pseudonyms (e.g., by requesting pseudonyms for the same period multiple times). This set the ground for Sybil-based misbehavior~\cite{douceur2002sybil}. The early proposal \cite{papadimitratos2007architecture} propose to equip each vehicle with a tamper-proof \ac{HSM}, which prevents adversaries from manipulating pseudonym acquisition and usage. However, in a heterogeneous system like \ac{VSN}, where various kinds of mobile devices are involved, such assumption would not stand.

% By extending the adversarial model, i.e., generalization of the system set up and considering honest-but-curious entities, a compromised entity is able to request to multiple \ac{PCA}, thus performing Sybil attack \cite{khodaei2014towards}.

It would be straightforward if the pseudonym provider simply issues pseudonyms with non-overlapping lifetimes and keeps a record of pseudonym issuance for each vehicle. Then, the pseudonym provider would know until when the issued pseudonyms are valid for a vehicle and only issue pseudonyms for the period after that. However, this would not work if there exist multiple pseudonym providers (so that a vehicle can request pseudonyms from the closest one) and a vehicle can request pseudonyms from different pseudonym providers. This is due to the separation of duties among authorities is enforced in \ac{VC} domain so that the pseudonym providers are not allowed to share their records (otherwise the authorities could collude and infer extra information) \cite{khodaei2014towards, gisdakis2014sppear}. \cite{khodaei2014towards} proposes a ticket-based scheme: an anonymized ticket is obtained from identity provider, then the ticket is used to obtain pseudonyms from a pseudonym provider. Each ticket is bound to a specific pseudonym provider without disclosing the targeted pseudonym provider to the identity provider, so that each ticket can be used only once while not revealing location of the vehicle to the identity provider. Moreover, the identity provider (only) learns for which period a vehicle has requested pseudonyms for, so no ticket will be issued again for the same period.

% It is not straightforward to enforce a policy on the identity provider to prevent this without revealing any useful information, e.g., actual pseudonym acquisition time. 

% Most of schemes proposed under this circumstances are vulnerable to Sybil-behavior. Detection mechanisms~\cite{viswanath2011analysis} can be utilized to detect Sybil node, however, the challenge here is to prevent such misbehavior from happening instead of detection, i.e., designing the \ac{VSN} infrastructures and system protocols to be resistance against Sybil-behavior. In~\cite{khodaei2014towards}, users are only allowed to obtain pseudonyms with non-overlapped lifetimes so that only one pseudonym is valid for each user at any point.

% a scheme \cite{puca2014} with zero-knowledge proof protocols \cite{camenisch2006win} as the \ac{VPKI} for the \ac{VC} system

% SOR \cite{luan2015social} uses a privacy preserving interest matching for social networking application on highways. It relies on choosing large prime numbers and convert the interest vector to a vector of prime numbers to be transmitted to another vehicle. In this case, the 

\subsection{Data Verification}

Entity authentication would help to eliminate illegitimate users and enhance the trustworthiness of the content being transmitted among the users. Leveraging post-misbehavior approaches (e.g., pseudonym resolution \cite{khodaei2014towards}), the entities that provided faulty information could be punished (e.g., revoked from the system). However, accepting false information (e.g., related to safety applications) from internal adversaries would be fatal and the problem would not be solved although the entities could be revoked from the system, if such misbehaviors have already led to accidents. This requires the received data to be verified and validated before it is accepted, i.e., the trustworthiness of the received data need to be verified even after the source and the integrity of the data is verified with signature verification.

\cite{dietzel2014redundancy, gisdakis2015shield} propose internal attacker detection approaches for sensing data aggregation based on redundancy of data received from multiple sensing entities, assuming the majority of the internal nodes are honest. In both works, the authors leverage entity authentication. In~\cite{dietzel2014redundancy}, each vehicle aggregates the received data which correspond to the same event and merge the data with its own sensing data and forward to neighboring vehicles. Each received message contains a path list and the redundancy is determined based on the nodes included in the path lists so that malicious nodes could not increase the redundancy of false information by affecting the aggregated data from multiple paths. In~\cite{gisdakis2015shield}, the sensing data from users' mobile devices are sent to and aggregated by the central server. The aggregated data then can be queried by the users.  The server detects the outliers purely based on the sensing data (e.g., temperature measurement) submitted from users' mobile devices. The server is trained with initial submissions and is updated with the successive submissions, then the outliers are detected based the training results.

\subsection{Fairness and Incentives}
\label{sec:fairness}

Collaboration among the nodes is the basis of security and privacy solutions in various domains (e.g., privacy-enhancing LBS~\cite{shokri2014hiding} and \ac{PS} networks \cite{burke2006participatory, ganti2011mobile}). Such solutions rely on participation and contribution from single entities to share workload of tasks or form groups to provide shelters for the nodes who wish to perform privacy-sensitive activities. In~\cite{shokri2014hiding}, \ac{LBS}-obtained information is shared with other peers so that the peers who is looking for the same information does not need to expose their location (and activities) to the possibly honest-but-curious \ac{LBS} servers. Crowdsourcing applications leverage contribution from the users in the system to infer context-dependent data (e.g., temperatures and humidity) when central providers do not exist. In principle, user experience would be improved with more user participation. However, without any guarantee for users' participation, selfish users could choose to benefit from the system while not contributing to the system (e.g., requesting information from neighbors while not sharing information with others). A motivation is needed for the user participation to keep the whole system operational.

Users' contribution can be monitored by the central infrastructures. \cite{lin2013achieving} proposes cooperative verification of safety beacons in \ac{VANET}, in which vehicles share the verification results with other vehicles so that each and every vehicle does not need to verify the signatures on all the received beacons. The whole process is monitored by the \acp{RSU} leveraging ID-based signcryption scheme~\cite{boneh2001identity,barreto2005efficient}. The \acp{RSU} provide each vehicle with a token for each time slot. A token is used by a vehicle to sign and encrypt its own verification effort (an integrated signature on multiple beacons), and decrypt and verify the verification efforts from other vehicles within the time slot. The vehicles have to prove to the \acp{RSU} that reasonable amount of effort have been made to obtain a new token for the next time slot, otherwise would not be able to decrypt the integrated signatures and benefit from other vehicles during the next time slot.

Incentivized scheme can be used to ensure the users' contribution, in which users are awarded virtual credits for their contribution. \cite{zhou2015secure, chen2010mobicent} propose incentive mechanism for data forwarding in \ac{DTN}, in which a central server stores the credits of different users. After each successful transmission, the credit is charged from the source node and distributed to intermediate nodes which relayed the packets. If a node does not actively participate in the transmissions, it would not have enough credits to send its own packets.

\section{Open Challenges} 
\label{sec:challenges}

Based on and beyond the existing solutions for security and privacy preserving \acp{VSN}, there still exist a number of significant security and privacy challenges towards deploying such \acp{VSN}. Next, we explain these challenges with the current (yet not complete) efforts towards them.

\subsection{Resilience Considerations}

Sybil resilience in \acp{VSN} remains an open challenge in the absence of consensus because the standardization bodies \cite{etsietsi1, 1609draft} and harmonization efforts do not have conclusive views on that front. For example, \ac{C2C-CC} \cite{c2c} proposes to issue pseudonyms with overlapping lifetimes in order to keep the safety applications operational at any given point in time \cite{khodaei2015VTMagazine}, while ~\cite{papadimitratos2007architecture, khodaei2014towards} (works in the context of SeVeCom~\cite{leinmueller2006sevecom} and PRESERVE EU project~\cite{stotz2011security} respectively) proposes to issue pseudonyms with non-overlapping lifetimes in order to eliminate the possibility of equipping a vehicle with multiple simultaneously valid short-term identities. However, beyond enforcing this constraint within a domain, malicious users could exploit the existence of multiple domains~\cite{alexiou2013vespa, gisdakis2013serosa, puca2014} to obtain simultaneously valid pseudonyms from different domains, depending on the pseudonym usage policies.

\cite{gisdakis2013serosa,papadimitratos2008secure} propose to use an \ac{HSM} module for pseudonym management and cryptographic operations in order to prevent malicious users from deviating system policies for the pseudonym usage. However, in a \ac{VSN}, a heterogeneous networks with various kinds of mobile devices involved, it is not realistic to assume that all the devices (e.g., smartphones) would be integrated with \acp{HSM}. Therefore, it is interesting to raise a question: \emph{how to prevent users, with no \ac{HSM} equipment, from obtaining pseudonyms from different \acp{PCA} in a multi-domain environment?} To the best of our knowledge, \cite{khodaei2014towards} is the only scheme that prevents Sybil-based misbehavior at the security infrastructure without relying on \ac{HSM}.

However, another issue still remains: \emph{how to prevent users from sharing pseudonyms with each other?} Users can still transfer the private keys and pseudonyms to other devices if a pseudonym is not explicitly bound to an entity. Even though the ownership of the pseudonym is maintained by the security infrastructure and the real identity of the owner of a pseudonym could be resolved in case of misbehavior, such an on-demand process does not support real-time detection and prevent the users from sharing/transferring their credentials in the presence of malicious users (e.g., two malicious nodes can share their pseudonyms and act as Sybil nodes in two different areas).

\subsection{Inference Attacks}

Open wireless networks inevitably face security and privacy issues because anyone can eavesdrop the messages and manipulate them. Messages eavesdropped in the \ac{VSN} can be used to infer user activities, profile users or track a specific user. An observer could leverage different techniques, e.g., data mining, with the publicly available information to link user messages and track them based on the geographical information included in the messages. In principle, inference attacks are feasible due to openness of the transmitted messages (i.e., user information is anonymized but message content is kept open). Keeping message content confidential would prevent external eavesdroppers, however, this would introduce extra processing delay and affect real-time operations in the system.

Due to the disclosed location information and the structure of the roads (i.e., mobility restrictions), it is not difficult to link users according to the available information. For example, an attacker who is able to eavesdrop all beacons within the \ac{VANET} is able to track the vehicles with almost 100\% accuracy \cite{wiedersheim2010privacy}. Meta-data in the disseminated data can be used as extra information to link users. For example, an adversary can link beacons based on, e.g., speed and direction, or even link pseudonyms based on their lifetimes~\cite{khodaei2014towards}. As the \ac{VSN} applications become popular, more and more data will be exchanged among the vehicles; thus, the privacy of users is at stake. 

\subsection{Operational Challenges in Identity Management} 

Facilitating cross domain operations is one of the main operational challenges in \ac{VC} systems \cite{khodaei2015VTMagazine}. Similar to \ac{VC} systems, trust establishment has to be taken into consideration before deploying a secure and privacy-preserving multi-domain \ac{VSN}. Trust establishment among the vehicles within the same domain, e.g., Volvo cars (assuming cars from the same manufacturer fall in a same domain), should be easy; however, in this case, it is not clear how a user with Volvo car should establish trust with a Toyota car. The key operational questions are: \emph{who will be operating the identity and credential provision?} More importantly, \emph{how the trust among the vehicles are established?} \cite{khodaei2015VTMagazine}.

On the other hand, pseudonymous authentication provides adequate level of security and privacy; however, revocation of the pseudonyms has not been fully addressed by academia and the industry. There are several challenges in pseudonym acquisition and revocation, e.g., connectivity to the \ac{CA} for fetching \ac{CRL}, acquisition of a large number of pseudonyms in an unstable network condition and the necessity to integrate a misbehavior detection mechanism to identify misbehaving entities and revocation of them.

% That is why in some proposals the revocation of the pseudonyms are ignored due to the mentioned practical constraints \cite{c2c}.

As described in \ref{sec:fairness}, there are the cases when user contribution needs to be recorded and incentivized: any contribution made under pseudonyms should be dedicated to corresponding long-term identities, otherwise the record would be lost once a new pseudonym is used. This could be achieved with the help of a central server keeping the credits of each user. However, such approach conflicts with the motivation behind pseudonymous authentication: any two pseudonyms of a same user should not be linkable, as it requires the central server to identify which pseudonyms corresponds to which long-term identities. This leaves a challenge of accumulating credits for user contribution while preserving user privacy.

% In group signature based systems, the misbehaving entity can be revoked by the group manager \cite{camenisch2006win}; however, an extensive performance validation of using these schemes is needed to evaluate to what extent these scheme can compete with classical public-key cryptography schemes \cite{khodaei2015VTMagazine}.

%Signing with a resolvable pseudonym is a preemptive countermeasure to prevent the clients from disseminating 

%*********************************************************************************

%Security and privacy solutions should be interoperable.  instead independent schemes for each application, otherwise increase complexity.

%Challenge:

%Confidentiality

%Privacy in incentive scheme

\section{Conclusions}
\label{sec:conclusions}

In this chapter, we surveyed and presented the state-of-the-art \ac{VC} systems, security and privacy architectures and technologies, emphasizing on security and privacy challenges and their solutions for P2P interactions in \acp{VSN} towards standardization and deployment. We note that beyond safety applications that have drawn a lot of attention in \ac{VC} systems, there is significant and rising interest in vehicle-to-vehicle interaction for a range of transportation efficiency and infotainment applications, notably \ac{LBS} as well as a gamut of services by mobile providers. While this enriches the \ac{VC} systems and the user experience, security and privacy concerns are also intensified. This is especially so, considering (i) the privacy risk from the exposure of the users to the service providers, and (ii) the security risk from the interaction with malicious or selfish and thus misbehaving users or infrastructure. We showed existing solutions can in fact evolve and address the \ac{VSN}-specific challenges, and improve or even accelerate the adoption of \ac{VSN} applications.

%\section{Glossary (maybe remove)}

%\printglossaries

%\cite{androulaki2008reputation} proposes a reputation-based system for anonymous networks using which each user is able to accumulate reputation while switching form one pseudonym to another one. Thus, a misbehavior cannot conceal his misbehaving activities with changing pseudonyms. In such an identity bound reputation system, a trust central entity, the bank, keep the identity of each user with their accumulated reputation points, i.e., the e-cash. There are also 

%\cite{krontiris2012monetary} monetary rewards 

%service quotas \cite{luo2012fairness}, 

\bibliographystyle{plain}
\bibliography{references,references1}

%\printindex

\end{document}